%..............................................................................
\magnification=1200 
\baselineskip=20pt
\def\rb{R_b}
\def\als{\alpha_s}
\def\gel{g_{E,L}}
\def\guer{g^U_{E,R}}
\def\gder{g^D_{E,R}}
\def\rc{R_c}
\def\rl{R_l}
\def\delrb{\delta R_b}
\def\tilT{{\tilde T}^c_L}
\def\Xmu{X_{s\mu}}
\def\gmmuu{\gamma^{\mu}}
\def\gmmud{\gamma_{\mu}}
\def\gblsm{g^b_{L, SM}}
\def\gbrsm{g^b_{R, SM}}
\def\delgbl{\delta g^b_L}
\def\delgbr{\delta g^b_R}
\def\ftc{F_{TC}}
\def\ntc{N_{TC}}
\def\delgmb{\delta\Gamma_b}
\def\gmb{\Gamma_b}
\def\delvbn{\delta^{new}_{vb}}
\def\delalsn{\delta\alpha^{new}_s}
\def\delgvan{\delta^{new}({g_{vl}\over g_{al}})}
\centerline{\bf Resolution of the $\rb -\als$ crisis of the SM}
\centerline{\bf in an exotic ETC scenario.}
\vskip .4truein 
\centerline{\bf Uma Mahanta}
\centerline{\bf Mehta Resarch Institute}
\centerline{\bf 10 Kasturba Gandhi Marg}
\centerline{\bf Allahabad-211002, India}
\vskip 1truein
\centerline{\bf Abstract}

In this article we show that the $\rb -\als $ crisis of the SM can be
resolved in a commuting TC scenario with sideways gauge bosons only if
the ETC representation contains mirror technifermions. For a toy ETC model
we estimate the value of $\xi^2_t={\gel/ \guer}$ that is needed to produce the
observed LEP excess in $\rb$. It is also shown that the same value of
$\xi^2_t$ resolvs the $\als$ crisis of the SM but produces a
$\delta \rho_{new}$ which is barely within the present experimental bound.
\vfill\eject

The SM works extremely well. All experimental data are quite well explained
in the context of the SM. However the precision electroweak measurements
at LEP have recently produced data about three observables $\rb, \rc and \als
(M_z)$ that seem to show some deviations from the SM predictions which
might well be the first hint of new physics beyond the SM. The first and most
widely reported deviation is the observed LEP excess in $\rb ={\Gamma
(z\rightarrow b{\bar b})\over \Gamma(z\rightarrow  h)}$.
Th average value of $\rb$ quoted by the LEP collaborations [1] is $\rb^{expt} 
\approx .2202 \pm .0020$. This is $2.4\sigma$ higher than the SM prediction
$\rb ^{SM}\approx .2156 \pm .0004$.  A somewhat closely related but less
prominent dicrepancy is the $1.3\sigma$ deficit in the LEP value of
$\rc ={\Gamma
(z\rightarrow c{\bar c})\over \Gamma(z\rightarrow  h)}$. The experimental
value [1] is $rc^{expt}\approx .1583\pm .0098$ which is to be compared with
the SM prediction $\rc^{SM}\approx .1711$. The third discrepancy concerns
the difference between the QCD coupling constant $\als (M_z)$ determined
from Z pole measurements and other low energy data [2]. Low energy
measurements [2] such as DIS and $\upsilon$ decay favor a value of 
$\als (M_z)$ close to 0.111. 
Lattice simulations of the bottomonium system [3] gives $\als (M_z)\approx
.115\pm .002$ which is consistent with the DIS value. On the other hand
high energy measurements at LEP based on $\rl={\Gamma(Z\rightarrow h)
\over\Gamma(Z\rightarrow l^+l^-)}$ gives the value [2,4] 
$\als (M_z)\approx .128\pm .005$. A global fit to all the Z line shape data,
including $\rl$, gives $\als (M_z)\approx .125\pm .004$. It is therefore
puzzling that the LEP values are $\als (M_z)$ are systematically higher
than the low energy values. Among these three deviations, the discrepancy
in $\rc$ is least serious first because the LEP deficit is only at the level
of $1.3 \sigma$ and second the charm fragmentation functions are not known
that accurately. 

Several ETC scenarios that can resolve the $\rb -\als$ crisis of the SM
have been proposed. In the traditional commuting ETC sceario, diagonal
ETC exchange tends to increase $\rb$ [5] whereas sideways ETC exchange
tends to decrease $\rb$ [6] relative to its SM value. On the other hand in
non-commuting ETC scenario [7] the sideways ETC induced vertex
 correction
tends to increase $\rb$ whereas the mixing between the two
 neutral Z bosons
tends to decrease it. The overall size and sign of ETC induced
 correction
$\delrb$ in both these scenarios is therefore model dependent and we
cannot make a definite prediction about $\rb$. In this article we will
consider a commuting ETC scenario whose fermionic representation
contains ordinary fermions with V-A weak interaction but TF's with
V+A weak interaction and we shall show that it can resolve the 
$\rb -\als$
crisis of the SM through sideways ETC exchange only. Unlike standard 
commuting ETC scenarios there is no need to invoke additional diagonal
ETC exchange to get a positive $\delrb$.

Usually in commuting ETC models LH ordinary fermions 
$(\psi_L=(t_L, b_L))$
and TF's $(T_L=(U_L, D_L))$ of $I_3=\pm 1/2$ are placed in identical
representations [8] of $G_{ETC}$. On the other hand to produce the
observed isospin breaking in the ordinary fermion mass spectrum
$(t_R, U_R)$ are placed in a different representation from
$(b_R, D_R)$. In this article we shall deviate somewhat from this 
requirement and assume that  $(\psi_L=(t_L, b_L))$ and
${\tilde T}^c_L=\imath \tau_2 T^c_L =(D^c_L, -U^c_L)$ with $I_3=\pm 1/2$
are placed in identical representations of  $G_{ETC}$. Here $T^c$ is the
charge conjugated TF doublet i.e. $T^C= C{\bar T}^T$.
 On the other hand 
$(t_R, D^c_R)$ with $I_3=1/2$ will be assumed to be placed in a different
representation from $(b_R, U^c_R)$ with $I_3=-1/2$ to produce the large
top-bottom mass splitting. Note that under $SU(2)_w\times U(1)_y$,
$\psi_L$ transforms as $(2, 1/6)$ and ${\tilde T}^c_L$ transforms as
(2, 0). On the other hand $U^c_R$ and $D^c_R$  trnansform as (1, -1/2)
and (1, 1/2) respectively under the same. 
 This implies that  TF's have V+A weak interaction. Also
 since $T^c_L=(U^c_L,
 D^c_L)$ transforms as $2^*$ under $SU(2)_w$ it cannot be placed 
 together with $\psi_L$ in the
same LH representation of $G_{ETC}$  for a commuting ETC scenario. 
It is clear from the above that the LH ETC representation does not commute
 with $U(1)_y$. The corresponding ETC gauge  boson $\Xmu$ therefore 
 carries
 hypercharge. $U(1)_y$ invariance implies that $\Xmu$ can
  mediate
 transition between $t_R$ and $D^c_R$ or between $b_R$ and  $U^c_R$
 but not between $t_R$ and $U^c_R$ or between $b_R$ and $D^c_R$.
 Following th usual practice of normalizing charged current
 interaction describing sideways intraction can be written as
 $$L_{ETC}=-{1\over {\sqrt 2}}(\Xmu J^{\mu}_s+h.c.).
 \eqno(1)$$
 where $J^{\mu}_s=\gel{\bar \psi}_L\gmmuu\tilT-\guer
 {\bar t_R}\gmmuu D^c_R+\gder{\bar b}_R\gmmuu U^c_R$. Here color
  and  technicolor indices have been suppressed. The above sideways 
 Lagrangian will give rise to the following masses for t and b:
 $$m_t\approx {\gel\guer <{\bar D} D>\over 2M_s^2}\ \ and \ \ m_b\approx
 {\gel\gder <{\bar U} U>\over 2M_s^2}.\eqno(2)$$
If the TC sector is isospin symmetric i.e. if $<{\bar U}U>=<{\bar D}D>$,
 the large heirarchy between $m_t$ and $m_b$ can be produced by ETC
 interactions provided $\guer\gg \gder$. Since we do not have any 
 realistic ETC model that produces this large heirarchy between $m_t$ and
 $m_b$ we have simply assumed different sideways couplings of $t_R$
 and $b_R$ to the same sideways ETC gauge boson. The product of LH(RH)
 sideways current with its h.c. gives rise to $Zb_L{\bar b}_L
(Zb_R{\bar b}_R)$ vertex correction. However since $\delta \Gamma_b
\propto 2[\gblsm\delgbl+\gbrsm\delgbr]$ and $\vert \gbrsm \vert\ll
\vert \gblsm \vert $ we can ignore the effect of $\delgbr $ on $\delrb $
provided $\vert \delgbr \vert\approx \vert \delgbl \vert $. 
The ETC induced correction $\delgbl $
can be derived from the following 4f sideways interaction
$$L^s_{4f}\approx -({\gel}^2/2 M_s^2){\bar \psi}_L\gmmuu\imath \tau_2
T^c_L{\bar T}^c_L(-\imath \tau_2)\gmmud\psi_L \ \ .\eqno(3)$$
Fierz transforming both w.r.t Dirac and $SU(2)_w$ indices and using the 
identities ${\bar T}^c_L\gmmuu T^c_L={\bar T}_R\gmmuu T_R, \ \ {\bar T}^c_L
\gmmuu\tau^*_i T^c_L={\bar T}_R\gmmuu \tau_i T_R $ we get
$$L^s_{4f}\approx -({\gel}^2/2 M_s^2) [{\bar \psi}_L\gmmud \psi_L 
{\bar T}_R\gmmuu T_R-{\bar \psi}_L\gmmud\tau_i\psi_L{\bar T}_R\gmmuu\tau_i
T_R].\eqno(4)$$
The sideways ETC induced vertex correction $\delgbl$ can be obtained from
the expression for $L^s_{4f}$ if we replace  the TF currents by the 
corresponding sigma model currents below the TC chiral symmetry breaking
scale [9]. We have ${\bar T}_R \gmmud T_R=(\imath F^2_{TC} /2)Tr[\Sigma 
(D_{\mu}\Sigma )^+]$ and ${\bar T}_R\gmmud\tau_i T_R =(\imath F^2_{TC} /2)
Tr[\Sigma\tau_i (D_{\mu}\Sigma )^+]$ where $\Sigma =e^{\imath \tau^i \pi_i
\over F_{TC}}$
and for mirror TF's
$$\eqalignno {D_{\mu}\Sigma & =\partial _{\mu}\Sigma
 -{\imath g\over 2{\sqrt 2}}\Sigma 
(W^+_{\mu}\tau^+ +W^-_{\mu}\tau^-)+\imath e[Q, \Sigma ]A_{\mu}\cr
&-{\imath e s_w\over c_w}[Q, \Sigma ]Z_{\mu}-{\imath g\over c_w}\Sigma
{\tau_3\over 2} Z_{\mu}.&(5)\cr}$$
In the unitary gauge $\Sigma =1$ and we get 
${\bar T}_R \gmmud T_R=0$ and
$${\bar T}_R\gmmud\tau_i T_R =-{F^2_{TC}\over 2}g
(W_{1\mu}\delta_{i1}+W_{2\mu}\delta_{i2})-{F^2_{TC}\over 2}(g/c_w)Z_{\mu}
\delta_{i3}.\eqno(6a)$$
The $Z\psi_L{\bar \psi}_L$ vertex correction is therefore given by
 $$\delta L_{z\psi_L{\bar\psi}_L}=-({\gel}^2/8M_s^2)F^2_{TC}
(g/c_w){\bar \psi}_L\gmmud\tau_3\psi_L Z^{\mu}\eqno(6b)$$.

In SM at tree level the $Zb_L{\bar b}_L$ coupling is given by
$L_{Zb_L{\bar b}_L}=-{g/c_w}\gblsm Z_{\mu}{\bar b}_L\gmmuu b_L $
where $\gblsm =-(1/2)+(1/3)s^2_w$. Since $\delgbl (sideways)=-{{\gel}^2
F^2_{TC}\over 8M_s^2}$ is of the same sign as $\gblsm$,
 the sideways ETC induced
correction in the mirror TF scenario tends to increase $\Gamma_b$ and 
hence $\rb$ (${\delrb\over \rb}=(1-\rb){\delta \Gamma_b\over \Gamma_b}$)
relative to its SM value. In standard ETC scenario the sideways 
exchange produces a negative $\delrb$ and one has to invoke additional
diagonal exchange to get an overall positive $\delrb$ to fit ythe LEP
data. Whereas
in our model the sideways exchange by itself produces a positive $\delrb$
and there is no need to invoke diagonal ETC exchange. The closure of
sideways currents however implies the existence of diagonal currents.
The diagonal ETC exchange will produce a negative $\delrb$ in our model.
The contribution of diagonal exchange can however be made negligible
by making the corresponding  gauge bosons much heavier than the sideways
gauge bosons. 
For simplicity we shall assume that the TF's are
placed in the fundamental representation of $SU(N)_{TC}$. Using naive
dimensional analysis [10] and large $N_{TC}$ scaling we can then write 
$<{\bar D}D>\approx 4\pi F^3_{TC}{\sqrt {3\over N_{TC}}}$.
From the xpression of $m_t$ (Eqno. 2) we get 
${{\gel}^2 {\ftc}^2\over M_s^2}\approx {m_t\over 2\pi \ftc}
{\gel\over \guer}
{\sqrt {\ntc\over 3}}\approx .1128 {\sqrt {\ntc\over 3}}
{\gel\over \guer}$ for $m_t\approx 175$ Gev and $\ftc\approx 247$ Gev.
Further using the relation ${\delrb\over \rb (1-\rb)}\approx {\delgmb\over
\gmb}\approx -4.5758 \delgbl $ we find that to produce the $2.4\sigma$
LEP excess in $\rb$, the ratio ${\gel\over \guer}$ must assume
 the values .517, .365, .299,  and .259 for $\ntc =$ 2, 4, 6 and 8
respectively. Hence if the sideways ETC scale is low enough to produce
an $m_t\approx 175 $ Gev, the values of ${\gel\over \guer}$ for $\ntc
\le 8$ that are required to produce the desired LEP excess in $\rb$
 are therefore quite natural and needs no fine tuning.

 The contribution of new phyics to $\delgmb$ also affects the determination
 of $\als (M_z)$ from $\rl$. Denoting the new physics contribution to
 $\gmb$, ${g_{vl}\over g_{al}}$ and $\als (M_z)$ by $\delvbn$, $\delgvan$
 and $\delalsn $ respectively we can write to a very good approximation
 [11]
$$\gmb \approx \gmb ^{SM} (1+\delvbn ).\eqno(7a)$$
$$ \delrb \approx {13\over 59} [{46\over 59}\delvbn + {24\over 767}
\delgvan +0.1 (\delalsn (M^2_z)/\pi)].\eqno(7b)$$
and
$$\delta \rl \approx {59\over 3} [{13\over 59}\delvbn +{20\over 59}
\delgvan +.328 \delalsn (M^2_z).\eqno(7c)$$
where $\delrb\approx .0046$ and $\delta \rl\approx .0340$. If we assume that
the new physics does not couple directly to leptons we have $\delgvan =0$.
From eqns. (7b) and (7c) we then get $\delvbn \approx .0277  $
 and $\delalsn (M^2_z) \approx -.0133 $. From the experimental value of 
 $\rl $ on can determine $\als (M^2_z)$ in the context of the SM in a
relatively clean way. The value obtained in this way  is $\als ^{SM}(M^2_z)
\approx .128 \pm .005 $. Incorporating the correction due to new physics
we get $\als^{new} (M^2_z) \approx .115 $ which is in excellent agreement
with the Lattice determination of $\als (M^2_z)\approx .115 \pm.002 $
from the $\upsilon$ system. It also agrres with the value of $\als (M^2_z)$
determined from DIS ($.113 \pm .005 $) within its uncertainties.

The low ETC scale that gives rise to the large mass for the top quark can
also produce observable weak isospin breaking effects [12]. In our model
there are two ETC induced  4f operators that lead to weak isospin breaking.
 $$O_1={{\gel}^2\over 4M_s^2}[{\bar \psi}_L\gmmud\tau_i\psi _L 
 {\bar T}_R\gmmuu\tau_i T_R -{\bar \psi}_L\gmmud\psi _L
 {\bar T}_R\gmmuu T_R ].\eqno(8a)$$
 and 
 $$O_2=-{(\guer )^2\over 2M_s^2}[{\bar D}_L\gmmud D_L {\bar t}_R
 \gmmuu t_R].\eqno(8b)$$
 In the above we have assumed that the product of two I=1 TF currents that
usually lead to the most dangerous weak isospin violation [12] in TC models
is subdominant. The reason being such potentially dangerous operators
arise from diagonal ETC exchange but the mass of such diagonal gauge
bosons have been assumed to be much greater than that of sideways 
gauge bosons.
 The terms in $O_1$ and $O_2$ that contain either
an isosinglet TF current or an isosinglet ordinary fermion current does
not contribute to $\Pi^{11}(q^2)$ or $\Pi^{33}(q^2)$ and therefore 
they can be dropped. Let ${\tilde \Pi}^{i,j}_{V,A}(q^2)$ and
${\bar \Pi}^{i,j}_{V,A}(q^2)$ denote the gauge boson self energy  
correction due to TF's and ordinary fermions respectively. It can then be
shown that the contribution of $O_1$ to $\Pi^{11}(0)-\Pi^{33}(0)$ is given
by $[\Pi^{11}(0)-\Pi^{33}(0)]_{O_1}=-{3\gel ^2 \ftc ^2 m_t^2\over 256\pi^2
M_s^2}$. To find the contribution of $O_2$ to   $\Pi^{11}(0)-\Pi^{33}(0)$
we shall make use of the fact that ${\tilde \Pi}^{33}_V(0)={\bar \Pi}^{33}
_V(0)=0$ due to exact conservation of neutral vector current. We then have
$[\Pi^{11}(0)-\Pi^{33}(0)]_{O_2}={3(\guer ) ^2 \ftc ^2 m_t^2\over 256\pi^2
M_s^2}\ln {\Lambda_s^2\over m_t^2}$ where $\Lambda_s^2\approx
{M_s^2\over \gel^2}$. Hence $\delta \rho_s \approx{3m_t^2\over 64 \pi^2
M_s^2} [(\guer )^2\ln {\Lambda_s^2\over m_t^2}-\gel^2] \approx .0031 $.
From a global fit of the LEP data one obtains the bound [1] $\delta\rho
_{new}^{expt}\le .4\%$. The sideways ETC induced correction $\delta \rho_s$
 is therefore barely consistent with the global fits to the LEP data.

 Although the mirror TF scenario presented here provides a better fit to the
 LEP data w.r.t $\delrb$ and $\als$, it does not produce the necessary
shift in $R_c$ to account for the observed LEP deficit. If the parametrs
of the ETC model are so chosen as to produce the observed LEP excess in
$\rb$, the corresponding shift in $R_c$ is given by $\delta R_c\approx
-R_c{\delrb\over 1- \rb}\approx -.0011$, which is only  8.59\% of the 
desired shift of -.0128. A toy grand unified model containing four ordinary
fermion families with V-A interaction and four TF families with V+A
 interaction can be
constructed based under the group $O'(14)=SO(14)\times K'$ where $K'(64_L)
=1$ and $K'({\bar 64}_R)=-1$. The irreducible spinorial representation
128 of O'(14) decomposes under $SO(10)\times SU(2)_H\times SU(2)_{TC}$ as
follows: $128=(16_L, 2, 1)\oplus ({\bar 16}_R, 2, 1)\oplus ({\bar 16}_L, 1, 2)
\oplus (16_R, 1, 2)$. 
\centerline{\bf References}
 
 \item{1.}  J. Erler and P. Langacker, UPR-0632T, hep-ph/9411203 (Feb 1996).
 
 \item{2.} T. Takeuchi, Fermilab-conf-95/173-T. (June 1995); M. Shifman,
 Mod. Phys. Lett. A 10, 605 (1995).
 
 \item{3.} A. X. El-Khadra, G. Hockney, A. S. Kronfeld and P. B. Mackenzie,
 Phys. Rev. Lett 69, 729 (1992); C. T. H. Davies, K. Hornbostel, G. P. 
 Lapage, A. Lidsey, J. Shigemitsu and J. Sloan, Phys. Lett. B 345, 42 (1995).
 
 \item{4.} I. Hinchliffe, LBL-36374, hep-ph/9501354 (Jan 1995).
 
 \item{5.} N. Kitazawa, Phys. Lett. B 313, 395 (1993); G. H. Wu Phys. Rev.
Lett. 74, 4137 (1995).

\item{6.} R. S. Chivukula, S. B. Selipsky and E. H. Simmons, Phys. Rev. Lett.
69, 575 (1992); R. S. Chivukula, E. Gates, E. H. Simmons and J. Terning,
Phys. Lett. B 311, 157 (1993).

\item{7.} R. S. Chivukula, E. H. Simmons and J. Terning, Phys. Lett. B
331, 383 (1994).

\item{8.} E. Eitchen and K. Lane, Phys. Lett. B 90, 125 (1980).

\item{9.} H. Georgi, Weak Interactions and Modern Particle Theory,
Benjamin-Cummings, Menlo Park, 1984.

\item{10.} A. Manohar and H. Georgi, Nucl. Phys. B 234, 189 (1984).

\item{11.} A. Blondel and C. Verzegnassi, Phys. Lett. B 311, 346 (1993).

\item{12.} T. Appelquist, M. J. Bowick, E. Cohler and A. Hauser, Phys.
Rev. D 31, 1676 (1985).
  
\end